\documentclass[aps,prl,twocolumn,groupedaddress]{revtex4-1}

\begin{document}


\title{A model for generating tunable clustering coefficients independent of the number of nodes in scale free and random networks }


\author{Vijay K Samalam}
\affiliation{Janelia Farm Research Campus, Howard Hughes Medical Institute, 19700  Helix Drive, Ashburn, VA 20147}


\date{\today}

\begin{abstract}
Probabilistic networks display a wide range of high average clustering coefficients independent of the number of nodes in the network.  In particular, the local clustering coefficient decreases with the degree of the subtending node in a complicated manner not explained by any current models.  While a number of hypotheses have been proposed to explain some of these observed properties, there are no solvable models that explain them all.  We propose a novel growth model for both random and scale free networks that is capable of predicting both tunable clustering coefficients independent of the network size, and the inverse relationship between the local clustering coefficient and node degree observed in most networks. 
\end{abstract}

\maketitle

Clustering coefficients measure the number of closed triangles that exist in a network.  They are important because they profoundly affect the dynamics of processes on probabilistic networks by providing local feedback loops.  They are also a mystery because there are currently no theoretical solvable models that explain all the properties associated with clustering observed in real networks\cite{newmanbookclustering, barabasibook}.  The most common idea proposed for some of the observed properties comes from social networks.  In a social network a person who is independently linked to two others is likely to introduce the two thus forming a triad closure or a triangle\cite{wattssmallworld}.  The plausibility of this idea has been explored by simulations and also by analysis of existing databases that represent scientific collaboration networks in physics and biology\cite{clusterideaholme, clusterideanewman}.  Other ideas include the presence of community structures in social networks, preferential attachment in scale free networks, and the presence of hierarchy to explain clustering\cite{communitiesnewman, hierarchybarabasi}.  However all theoretical models proposed to date predict that the average clustering coefficient decreases to zero with the network size, $\mathcal{N}$, going as $1/\mathcal{N}$ for random networks and going as $((ln\mathcal{N})^2)/\mathcal{N}$ for scale free networks\cite{clusterscalefree}.  But analysis of all existing probabilistic networks shows that the average clustering coefficient is independent of the number of nodes and takes on a wide range of values.  Moreover none of the theoretical models to date seem capable of explaining the complex inverse dependence of the local clustering coefficient on the node degree.  In this work we propose a simple model for the growth of both random and scale free networks where the average clustering coefficient is tunable and independent of network size, and the local clustering coefficients node degree dependence can be specified.  Moreover, our analysis suggests that high clustering coefficients may not be tied either to the scale free nature of some networks nor to their possible hierarchical structure.
\par
The model we propose is again inspired by the old idea of how social networks grow, where when person A is introduced to person B, there is a higher probability than chance, that  person B will introduce A to their friends.  Moreover, if B has a lot of friends she might not introduce A to all of them but only some fraction of them.  While this model is a modification of the original Price model for the growth of directed network  it can be easily extended to the growth of non directed networks like the Albert Barabasi model and the main conclusions do not change\cite{Price:1976, Barabasi:1999}. At $t=0$, the network has $n_o$ nodes, all of them with zero in-degree links.  At every time step $t+1$, a new node with $(m+r)$ out-degree links attached to it joins the network. The $(m+r)$ links attach themselves to the other nodes in two steps. At time $t+\frac{1}{2}$, the ends of the m links attach themselves to the existing nodes following the usual linear preferential attachment rule where the probability of attachment is proportional to $(a+k)$, where $a>0$, ensures that nodes in the beginning with zero in-degrees get links, and $k$ is the in-degree of the node to which one of the m links attaches itself.  After the $m$ links attach themselves, the $r$ links attach themselves to the rest of the nodes using a different probability rule. The r links attach themselves with a higher probability to the first neighbors of the nodes to which the m links attached themselves and with a lower probability to all other nodes in the network.  It is some of these $r$ links that form the triangles in the network.  At time $t$, the network has $\mathcal{N} = n_0 + t$ nodes and $(m+r)t$ links.  Before calculating the clustering coefficients, we want to calculate the degree distribution of the network for large $t$ and especially for large $k$.  This is important since we want to make sure that a lot of desirable properties associated with random networks and scale free networks, like the small mean distances and diameters of the network, are maintained for this model. We derive a rate equation for changes to $N(k,t)$, the averaged number of nodes with in-degree $k$ at time $t$ and solve for it in the large $t$ and $k$ limit closely following some of the techniques used in the past for growing scale free networks\cite{Krapivsky:2000, DMS:2000}.  To do that we first calculate the expression for the change in $N(k,t)$ after time $t+1$. Since the probability of one of the $m$ links attaching themselves to a network node goes as
\begin{equation}
\frac{a+k}{\sum\limits_{i=1}^\mathcal{N}\left(a+k_i\right)} = \frac{a+k}{an_o + t\left(a+m+r\right)},
\end{equation}
 we have at time $\left(t+\frac{1}{2}\right)$
\begin{widetext}
\begin{equation}
N\left(k,t+1/2\right) = N\left(k,t\right)\left[1 - \frac{m\left(a+k\right)}{an_o+t\left(a+m+r\right)}  \right] + \frac{N\left(k-1,t\right)m\left(a+k-1\right)}{an_o + t\left(a+m+r\right)}.
\end{equation}
\end{widetext}
At this time the arriving node has acquired $\sum\limits_{i=1}^m k_i$ new second neighbors, where $k_i$ is the degree of one of the $m$ nodes to which the arriving node has attached itself. In the next stage of this process, the $r$ links now attach themselves following a different set of rules.  If $p(k_i)$ is the probability of one of the $r$ links attaching themselves to any of the new second neighbors and $q$ is the probability of them attaching to all the other nodes in the network that are not second neighbors then the probabilities satisfy the equation 
\begin{equation}
\sum\limits_{i=1}^m k_i p\left(k_i\right) +q\left[\mathcal{N}\left(t\right)-\sum\limits_{i=1}^m k_i\right].
\end{equation}
Notice that we have assumed that $p$ can in general depend on the degree of the node $k_i$ that is subtending the triangles.  If the probabilities for attaching one of the $r$ links to the nodes were a constant then we would have $p=q=1/\mathcal{N}\left(t\right)$.  We do not need a detailed form for $p\left(k\right)$ to continue the analysis though it is clear that it must be of the order of $\mathcal{N}$ if we want the $r$ links to preferentially attach themselves to their new neighbors. 
At the end of time $t+1$, the net decrease in $N(k,t)$ because of the attachment of $m+r$ links, for $k>0$, is given by
\begin{widetext}
\begin{equation}
-\Delta N\left(k,t\right) = \frac{m\left(a+k\right)N\left(k,t\right)}{an_o + t\left(a+m+r\right)} + rq\lbrace N\left(k,t+\frac{1}{2}\right) - \sum\limits_{i=1}^m k_i\rbrace + r\sum\limits_{i=1}^m p\left(k_i\right)N_{k_i}^1\left(k,t\right).
\end{equation}
\end{widetext}
The first term on the right hand side is the decrease because of the attachment of the $m$ links to $k$ degree nodes, the second term describes the attachment of the $r$ links to any $k$ degree node except those that are second neighbors of the incoming node, and finally the third term describes the formation of triangles by the $r$ links when they attach themselves to second neighbors of the incoming node. Here $N^1_{k_i}\left(k,t\right)$ is the number of $k$ degree nodes that are first neighbors of the $i^{th}$ node (one of the $m$ nodes that the arriving node has attached itself to). 
Since we are only interested in stationary solutions for the $k$ degree distribution in the $t\longrightarrow\infty$ limit and in that limit, $N\left(k,t\right)\longrightarrow P\left(k\right)\left(n_o+t\right)$, the terms 
\begin{widetext}
\begin{equation}
\frac{m \left(a+k\right)}{an_o +t\left(a+m+r\right)},\quad\frac{\sum\limits_{i=1}^m k_i}{N\left(k,t\right)} < \frac{mk_{max}}{N\left(k,t\right)},\quad \frac{\sum\limits_{i=1}^m N^1_{k_i}\left(k,t\right)}{N\left(k,t\right)} < \frac{m k_{max}}{N\left(k,t\right)}
\end{equation}
\end{widetext}
 are much less than one and all tend to zero. 
In addition, in this limit when the distribution is stationary
\begin{equation}
 \frac{N^1_{k_i}\left(k,t\right)}{k_i}  \approx \frac{N\left(k,t\right)}{n_o+t}.
\end{equation}
With these approximations the expression for the decrease in $N\left(k,t\right)$ simplifies to 
\begin{widetext}
\begin{equation}
-\Delta N\left(k,t\right) = \frac{m\left(a+k\right)N\left(k,t\right)}{an_o+t\left(a+m+r\right)} + \frac{rN\left(k,t\right)}{n_o+t}.
\end{equation}
\end{widetext}
 There is a similar term for the increase in the mean number of $k$ degree nodes where $k$ is replaced by $k-1$ in the above formula.  Putting this together gives for the rate equation in the continuous time approximation
 \begin{widetext} 
\begin{equation}
 \frac{dN\left(k,t\right)}{dt} + \frac{m\left(a+k\right)N\left(k,t\right)}{an_o+t\left(a+m+r\right)} + \frac{r}{n_o+t} = \frac{m\left(a+k-1\right)N\left(k-1,t\right)}{an_o+t\left(a+m+r\right)} + \frac{rN\left(k-1,t\right)}{n_o+t},
\end{equation}
\end{widetext}
with the initial condition for the equation taking the form $N\left(k,0\right) = 0, k>0$.
 The equation for $N\left(0,t\right)$ is different because we introduce a new zero degree node at every time step and the equation can be easily shown to be
\begin{equation}
 \frac{dN\left(0,t\right)}{dt} + \frac{mN\left(0,t\right)a}{an_o +t\left(a+m+r\right)} + \frac{rN\left(0,t\right)}{n_o+t} = 1, 
\end{equation}
with the initial condition being $N\left(0,0\right) = 0$.
 It is interesting to note that the above equations represent a growth model where $m$ links attach themselves preferentially to nodes with high $k$ degree, while the $r$ links attach themselves randomly to all the network nodes.  The possible increase to the number of degree $k$ nodes that might result from the formation of all those triangles is made up for by a decrease by the same number when the $r$ links attach themselves to non-triangle producing nodes.  As far as the degree distribution is concerned, to a very good approximation, the formation of triangles makes no difference. The solution to the above equations follows the standard procedure for solving these equations.  The stationary solution for the degree distribution then takes the form 
\begin{equation}
 P\left(k\right) = \lim_{t\to\infty} \frac{F\left(k,t\right)\left(an_o+t\left(a+m+r\right)\right)}{n_o+t}\lbrace1+ O\left(\frac{1}{t^2}\right)\rbrace,
 \end{equation}
 where the function $F\left(k\right)$ can be evaluated iteratively.  The final stationary solution then takes the expected form
 \begin{equation}
 P\left(k\right) = \frac{B\left[\left(a+r/\widetilde{m}+k\right), \left(1+\frac{1}{\widetilde{m}}\right)\right]}{B\left[\left(a+\frac{r}{\widetilde{m}}\right),\frac{1}{\widetilde{m}}\right]}\qquad\mbox{where},
 \end{equation}
 $B\left[x,y\right]$ is the incomplete Beta function and $\widetilde{m} = m/\left(a+m+r\right)$.
 In the large $k$ limit, the degree distribution decays as a power law with exponent given by $2+\left(a+r\right)/m$. Since $r$ can be absorbed into $a$, it is clear that to a very good approximation, the extra $r$ links in the growth model do not change the distribution in any meaningful way.  For large $a$ the degree distribution decays exponentially with $k$, reflecting the fact that the incoming links are now being assigned randomly to the network nodes thus providing a growing model of a random network. 
\par
 Now that it is clear that the model is capable of re-producing the degree distribution of either a random or scale free network we can calculate the clustering coefficient.  If $T\left(k,t\right)$ is the total number of triangles centered on all nodes of degree $k$ at time $t$, it can be shown that
 \begin{widetext}
 \begin{equation}
 \frac{dT\left(k,t\right)}{dt} =  \frac{T\left(k,t\right)}{N\left(k,t\right)} \frac{dN\left(k,t\right)}{dt} + \frac{mr\left(a+k-1\right)N\left(k-1,t\right)}{an_o +t\left(a+m+r\right)} \left(k-1\right)p\left(k\right).
 \end{equation}
 \end{widetext}
 The first term on the right hand side gives the change in the number of triangles centered on nodes of degree $k$ because the total number of degree $k$ nodes has changed because of the combined effect of the $m$ links attaching themselves preferentially and some of the $r$ links attaching themselves randomly to degree $k$ and degree $k-1$ nodes.  The second term on the right hand side gives the new triangles that are formed when some of the $r$ links attach themselves preferentially to their new neighbors formed by the attachment of some of the $m$ links to $k-1$ degree nodes.  We again search for stationary solutions for the average number of triangles per degree $k$ node in the large $t$ limit,
 \begin{equation}
 \widetilde{T}\left(k\right) = \lim_{t\to\infty} \frac{T\left(k,t\right)}{N\left(k,t\right)}.
\end{equation}
Using the above form for $T\left(k,t\right)$ and the asymptotic forms for $N\left(k,t\right)$ derived before we get
\begin{equation}
\widetilde{T}\left(k\right) = \frac{\widetilde{m}r\left(k-1\right)\left(a+k-1\right)p\left(k\right)P\left(k-1\right)}{1-P\left(k\right)}.
\end{equation} 
Defining the local clustering coefficient as 
\begin{equation}
C_{local}\left(k\right) = \frac{2\widetilde{T}\left(k\right)}{k\left(k-1\right)},
\end{equation}
 we get 
 \begin{equation}
 C_{local}\left(k\right) = 2\widetilde{m}rf\left(k\right).
 \end{equation}
 The global average clustering, $C$ coefficient is then  the local clustering coefficient averaged over the degree distribution which gives 
\begin{equation}
C = 2\widetilde{m}r \overline{f\left(k\right)}
\end{equation}
a form for the clustering coefficient independent of the number of nodes in the network. 
Notice that $m$ can be fixed by the average degree of the network, while $a$ can be tuned to fix the exponent of the large $k$ dependence of scale free networks and  $r$ and $p\left(k\right)$ can be used to fix the $k$ dependence of the local clustering coefficient and the value of the global average value. 
\par
In conclusion, we have presented a simple growth model of a random or scale free network where the average clustering coefficient is tunable and independent of the number of nodes in the network.  We have also shown that the $k$ dependence of the local clustering coefficient can be specified to match the actual function observed in real networks. While there have been suggestions in the literature that the large clustering coefficients are somehow due to some as yet not understood process in the growth of scale free networks, or due to the hierarchical structure observed in many networks, our model suggests that the simple idea of a preferential attachment of links to the neighbors of a just attached node might be sufficient to explain the large value of clustering coefficients observed. 
\section{}
\subsection{}
\subsubsection{}

\bibliography{networks}

\end{document}